\documentclass[12pt]{revtex4}
\usepackage{dcolumn}
\usepackage{graphicx}
\usepackage{amsmath}
\usepackage{amsfonts}
\usepackage{amssymb}
\usepackage{psfrag}
\usepackage{wrapfig}
\usepackage{subfigure}
\usepackage{makeidx}
\usepackage{bm}
\usepackage{epsf}
\usepackage{hyperref}

\begin{document}

\title{Spherically symmetric gravitational collapse of a dust cloud in
Einstein-Gauss-Bonnet Gravity}

\author{ Kang Zhou$^{1}$, Zhan-Ying Yang$^{1}$\footnote{Email:zyyang@nwu.edu.cn},
De-Cheng Zou$^{1}$ and Rui-Hong Yue$^{2}$\footnote{ Email:yueruihong@nbu.edu.cn}}

\affiliation{ $^{1}$Department of Physics, Northwest University, Xi'an, 710069, China\\
$^{2}$Faculty of Science, Ningbo University, Ningbo 315211, China}
\date{\today}

\begin{abstract}
\indent

We explore the gravitational collapse of a spherically symmetric dust
cloud in the Einstein-Gauss-Bonnet gravity without a cosmological constant, and obtain three
families of LTB-like solutions.
It is shown  that the Gauss-Bonnet term has a profound influence on the
nature of  singularities, and the global structure of space-time
changes drastically from the analogous general relativistic case.
Interestingly, the formation of a naked, massive and uncentral singularity,
 allowed in  5-dimensional space-time, is forbidden if $D\geq6$.
Moreover, such singularity is gravitational
strong and  a serious counter example to CCH.
\end{abstract}



\maketitle

\section{Introduction\label{intro}}
\indent

The seminal work of Oppenheimer and Snyder led to the establishment
viewpoint that the end state of spherically symmetric gravitational
collapse is a black hole which have an event horizon to external
observers \cite{Oppenheimer:1939ue}.
Hawking and Penrose proven that, according to
general relativity, there must be a singularity of infinite density
of matter and space-time curvature formed by gravitational collapse
\cite{Penrose:1964wq, Hawking:1969vz, Hawking:1973uf}. At the
singularity, the laws of science and the ability to predict the
future would break down. This led Penrose to propose the (weak and
strong) cosmic censorship hypothesis (CCH) \cite{Penrose:1969},
which to date remain unproven. The weak CCH asserts that there can
be no singularity visible from future null infinity, light from
singularity is completely blocked by the event horizon,
while the strong CCH prohibits singularity's visibility by any
observer. Despite almost 40 years of effort we still don't have a
general proof of CCH \cite{Joshi:2000fk}. On the contrary, it was shown
that the work of Oppenheimer and Snyder
\cite{Oppenheimer:1939ue} which forbidden the formation of the naked singularity
is not a typical model, and the
singularities formed in generic collapse are naked and observable
\cite{Christodoulou:1984mz, Nolan:2002mz}. Recently, Virbhadra, Ellis, Keeton, etc showed that black holes and naked singularities can be observationally differentiated through
their gravitational lensing effects. They classified naked singularities into 3 categories: weakly naked, marginally strongly naked, strongly naked  singularities \cite{Ellis:2002mz, Virbhadra:2008mz}, and hypothesize a new cosmic censorship: Generically, marginally and strongly
naked singularities do not occur in a realistic gravitational collapse \cite{Virbhadra:2009mz}. They conjecture that in case the weak CCH of Penrose turns out to be incorrect, the new cosmic censorship will hold good.

In recent years, superstring/M theory caused a renewed interest about higher order
gravity in more than $4$ dimensions. The most extensively studied model in
the effective low-energy action of superstring theory \cite{Lust:1989tj} is
so-called Einstein-Gauss-Bonnet
gravity (EGB) \cite{Lovelock:1971yv}. An interesting question will be, how
does Gauss-Bonnet term change the collapse of a massive body? Recently, Maeda considered the spherically symmetric gravitational
collapse of a inhomogeneous dust with the $D \ge 5$-dimensional
action including the Gauss-Bonnet term without finding the
explicit form of the solution \cite{Maeda:2006pm}. Then, Jhingan and
Ghosh presented a exact model of the gravitational collapse of a
inhomogeneous dust in five dimensional EGB \cite{Jhingan:2010zz, Ghosh:2010jm}. The
objective of this paper is to extend their work to a more general
case. We investigate the spherically symmetric gravitational collapse of
an incoherent dust cloud by considering a Lemaitre-Tolman-Bondi type
space-time in EGB without the cosmological constant,
and give three families of LTB-like solutions, which are hyperbolic,
parabolic and elliptic, respectively.
In order to find the final fate of the collapse, we discuss the formation
of singularities, and analyze the nature of them. In particular,
we consider that whether such singularities would be hidden or be
visible to outside observers. We also investigate the
gravitational strength of naked singularities, and discuss that
whether the naked singularities in EGB theory is a serious threat to CCH.

This paper is organized as follows. In Sec \ref{22s}, we give the
field equations in EGB without a
cosmological constant, and derive the LTB-like solutions. In
Sec \ref{33s}, we investigate the formation of the singularity of the
spherically symmetric gravitational collapse, and analyze the global structure of space-time. The
strength of the singularity is demonstrated in Sec \ref{55s}. Sec
\ref{66s} is devoted to conclusions and discussions.

Throughout this paper we use units such that $8\pi G = {c^4} = 1$.

\section{Equations and solutions \label{22s}}
\indent

The action of EGB can be written as
\begin{eqnarray}
S=\int{{d^D}x\sqrt{-g}(R+{\alpha}{\cal L}_2)+{S_{matter}}},\label{eq:1a}
\end{eqnarray}
where the coupling constant $\alpha$
can be regarded as the inverse of string
tension and be assumed $\alpha>0$ in this paper. For future simplicity,
we take the coefficient $\alpha=\frac{\tilde{\alpha} }{(D-4)(D-3)}$.
The second order (Gauss-Bonnet) term of Lovelock Lagrangian is given by
\begin{eqnarray}
{\cal L}_2 =R_{abcd}R^{abcd}-4R_{ab}R^{ab}+{R^2}.\nonumber
\end{eqnarray}
Varying the action Eq.~(\ref{eq:1a}), we obtain the field equations as
\begin{eqnarray}
{G_{ab}}& =& G_{ab}^{( 1 )} + {\alpha}G_{ab}^{( 2 )}=
{T_{ab}},\label{eq:2a}\\
G_{ab}^{(1)}&=&{R_{ab}} - \frac{1}{2}R{g_{ab}}, \nonumber\\
G_{ab}^{(2)}&=&2( - {R_{acde}}R_{~~~b}^{dec} - 2{R_{acbd}}{R^{cd}} -
2{R_{ac}}R_{~b}^{c} + R{R_{ab}}) - \frac{1}{2}{\cal L}_2{g_{ab}}.\nonumber
\end{eqnarray}

Assume that the system consists of a freely falling perfect fluid whose
energy-momentum tensor in comoving frame is \begin{eqnarray}
T_{\mu\nu}=\varepsilon (t,r)u_\mu u_\nu,\label{eq:3a}
\end{eqnarray}
where $u_\mu=\delta_\mu^t$ is the D dimensional velocity vector field.
The metric takes the form
\begin{eqnarray}
ds^2 =  - dt^2 + A( {t,r} )^2dr^2 + R( {t,r} )^2d\Omega
_{D-2}^2,\label{eq:4a}
\end{eqnarray}
where the coordinate $r$ is the comoving radial coordinate, $t$
the proper time of freely falling shells. Both $R$  and $A$ are
 un-negative function of $t$ and $r$.
As well known, in general relativity, such metric leads to the
Lemaitre-Tolman-Bondi (LTB) solution \cite{Lemaitre:1933gd}, which
has been extensively used not only in spherical collapse, but also
in cosmology as well \cite{Maartens:1995dx, Garfinkle:2006sb}.
Plugging the metric Eq.~(\ref{eq:4a}) into Eq.~(\ref{eq:2a}), we
find that
\begin{eqnarray}
G_r^t=\frac{(D -
2)(\dot{A}R'-A\dot{R}')\left\{2\tilde{\alpha}[A^2(1+\dot{R}^2)-R'^2]+
A^2R^2\right\}}{A^5R^3}=0,\label{eq:5a}
\end{eqnarray}
where an over-dot and prime denote the partial derivative with
respect to $t$ and $r$, respectively. Eq.~(\ref{eq:5a}) leads to two
families of solutions
\begin{eqnarray}
A(t,r)=\frac{R'}{W}\label{eq:6a}
\end{eqnarray}
and
\begin{eqnarray}
A(t,r) = \pm \frac{\sqrt {2\tilde{\alpha}}R'}{[R^2+ 2\tilde{\alpha}(
\dot{R}^2 + 1)]^{\frac{1}{2}}},\label{eq:7a}
\end{eqnarray}
where $W=W(r)$ is an arbitrary function of $r$. since Eq.~(\ref{eq:7a}) doesn't recover
the solution in Einstein gravity in the limit $\alpha \to 0$,  we shall only
consider the case $A =\frac{R'}{W}$ here.
The freedom of  re-labeling  spherical dust
shell by $r \to g( r )$ enables us to  fix
the hypersurface such that at $t = 0$, $r$ coincide with the area radius
\begin{eqnarray}
R( {0,r} ) = r.\label{eq:8a}
\end{eqnarray}

It is clear that  the equations of momentum conservation $(
T_i^\mu)_{;\mu }=0$ are automatically satisfied, and the $t$-component
 reads
\begin{eqnarray}
- \frac{\partial \varepsilon}{\partial t}-\varepsilon(
\frac{(R'^2)_{,t}}{2R'^2}+\frac{(D-2)(R^2)_{,t}}{2R^2})
=\frac{\partial }{{\partial t}}( {\varepsilon {R^{D-2}}R'})=0.\nonumber
\end{eqnarray}
Thus we have
\begin{eqnarray}
\varepsilon (t,r)&=&\frac{{\varepsilon(0,r){r^{D-2}}}}{R^{D-2}R'},\nonumber
\end{eqnarray}
and the mass function is
\begin{eqnarray}
M(r)=\frac{2}{D - 2}\int \varepsilon(t,r)R^{D - 2}d{R}=\frac{2}{D
- 2}\int_0^r \varepsilon(0,r)r^{D - 2}dr,\label{eq:9a}
\end{eqnarray}
which is positive and increases with increasing $r$ for the
non-negative energy density $\varepsilon (t,r)$.

According to Eq.~(\ref{eq:6a}), other field equations can be written as
\begin{eqnarray}
G_t^t&=&-\frac{(D-2)[\tilde{\alpha}R^{D-5}(\dot{R}^2+1-W^2)^2+
R^{D-3}(\dot{R}^2+1-W^2)]'}{2R^{D-2}R'}=-\varepsilon(t,r)\label{eq:10a}\\
G_r^r&=&-\frac{(D-2)}{2R^4}\left\{2(D-5)\tilde{\alpha}(\dot{R}^2+1-W^2)^2\right.\nonumber\\
&+& \left.[4\tilde{\alpha}R\ddot{R}+(D-3)R^2](\dot{R}^2
+1-W^2)+2\ddot{R}R^3\right\}=0\label{eq:11a}\\
G_{\theta_n}^{\theta_n}&=&\frac{(R^3G_r^r)'}{(D-3)R^2R'}=0.\label{eq:12a}
\end{eqnarray}
With the help of Eq.~(\ref{eq:9a}), Eq.~(\ref{eq:10a}) reduces to
\begin{eqnarray}
\tilde{\alpha}(\dot{R}^2+1-W^2)^2+R^2(\dot{R}^2+1-W^2)-\frac{M(r)}{R^{D
- 5}}=0,\nonumber
\end{eqnarray}
namely
\begin{eqnarray}
\dot{R}^2=\frac{-R^2+\sqrt{R^4+4\tilde{\alpha}\frac{M(r)}{R^{D-5}}}}
{2\tilde{\alpha}}+W^2-1.\label{eq:13a}
\end{eqnarray}
It is straightforward to check that  the  Eq.~(\ref{eq:13a}) ensures
Eq.~(\ref{eq:11a}) and Eq.~(\ref{eq:12a}) to be satisfied. Generally, the solution of
Eq.~(\ref{eq:13a}) can be classified into three types hyperbolic, parabolic and
elliptic corresponding $W(r) > 1$, $W(r) = 1$ or $W(r) < 1$
respectively. The case $W(r)=1$ is
a marginally bound case in which the metric Eq.~(\ref{eq:4a}) takes
the form of Minkowski metric on the hypersurface $t=0$. In the
present discussion we shall assume that $W(r)^2 >
1-\frac{-R^2+\sqrt{R^4+4\tilde{\alpha}\frac{M(r)}{R^{D-5}}}}
{2\tilde{\alpha}}$, which is equivalent to that $\dot{R}^2>0$.
For gravitational collapse,  $R(t,r)$ shall decrease
respecting to $t$, it gives
\begin{eqnarray}
\dot{R}=-\sqrt{\frac{-R^2+\sqrt{R^4+4\tilde{\alpha}\frac{M(r)}{R^{D-5}}}}
{2\tilde{\alpha}}+W^2-1}.\label{eq:14a}
\end{eqnarray}

For arbitrary initial data of energy density $\varepsilon(0,r)$,
Eq.~(\ref{eq:14a}) completely specify the dynamical evolution of
collapsing dust shells. In the general relativistic limit $\alpha
\to 0$ with marginally bound case, Eq.~(\ref{eq:14a}) can be
integrated to yield
\begin{eqnarray}
t_b(r)-t(r)=\frac{2}{D-1}\frac{R^{\frac{D-1}{2}}}{\sqrt{M(r)}},\nonumber
\end{eqnarray}
where $t_b(r)$ is an function of integration and
equals to $\frac{2}{D-1}\frac{r^{\frac{D-1}{2}}}{\sqrt{M(r)}}$
by using Eq.~(\ref{eq:8a}). Notice that the expression for
function $R(t,r)$ is similar to LTB and LTB-like (in the space-time with more than $4$ dimensions)
solutions in general relativity \cite{Lemaitre:1933gd, Ghosh:2001fb}.

Now, we return to discuss the above LTB-like solution  attached at the boundary of a dust cloud to the outside vacuum region.
The boundary   is represented by a finite constant comoving radius $r=r_0>0$, and the outside vacuum is
\begin{eqnarray}
ds^2=-F(r')dT^2+\frac{dr'^2}{F(r')}+r'^2d\Omega_{D-2}^2,\label{eq:00a}
\end{eqnarray}
with
\begin{eqnarray}
F(r')=1+\frac{r'^2}{2\tilde{\alpha}}[1-(1+\frac{8\tilde{\alpha} m}{(D-2)\Omega_{D-2}r'^{D-1}})^{\frac{1}{2}}].\nonumber
\end{eqnarray}
If we define that
\begin{eqnarray}
m&=&\frac{(D-2)\Omega_{D-2}M}{2},\nonumber\\
R&=&r',\nonumber\\
dT&=&\frac{W(r)}{F(R)}dt,\label{eq:020a}
\end{eqnarray}
and follow Maeda's method in \cite{Maeda:2006pm}, we find that in present case,
the LTB-like solution attached at the hypersurface $r=r_0$ to the outside vacuum solution smoothly.

\section{The Final fate of  collapse\label{33s}}
\indent

First, we consider the
formation of the singularity. There are two kinds of
singularities which are shell crossing singularities and shell
focusing singularities, defined by $R' = 0$ and $R = 0$,
respectively. The characteristic of a singularity in space-time
manifold is the divergence of the Riemann tensor and the energy
density \cite{Hawking:1973uf}. In our case, the Kretschmann
invariant scalar ${\cal K} = R_{abcd}R^{abcd}$ for the
metric Eq.~(\ref{eq:4a}) is
\begin{eqnarray}
{\cal K} = \frac{2[ \ddot{R}'^2R^4 + 2(D-2)\ddot{R}^2R'^2R^2 +
2(D-2)R^2\dot{R}^2\dot{R}'^2 + (D-2)(D-3)R'^2\dot{R}^4 ]}{R'^2R^4},\label{eq:15a}
\end{eqnarray}
and  can be shown to be finite on the
initial data surface. The  energy density of the fluid
dust sphere is $\varepsilon (t,r) = \frac{(D-2)M'}{2R^{D-2}R'}$.
It is clear that both Kretschmann scalar and energy density
diverge when $R' = 0$ and $R = 0$. Hence, we have the shell
crossing  and shell focusing singularities.
Shell crossing singularities can be naked, and has been shown to be gravitationally weak in
the LTB case, which  can not be consider seriously in the context of the CCH
\cite{Newman:1985gt}.
On the other hand, in general relativity, shell focusing
singularities can also be naked and gravitationally strong as well.
 Henceforth, shell focusing singularities are considered to be the
genuine singularities in space-time manifold
\cite{Christodoulou:1984mz, Deshingkar:1998cb}. Here we only limit
in the shell focusing singularities.

It is known that the gravitational collapse in the 5D EGB with the marginally
bound case $W(r)=1$ leads to a shell focusing singularity, and positive
$\tilde{\alpha}$ delays the formation of the singularity \cite{Jhingan:2010zz}.
In higher dimensional EGB, one can easily find
\begin{eqnarray}
\dot{R} < 0, \quad
\ddot{R}=\frac{R^3-(D-5)\tilde{\alpha}R^{4-D}M}{2\tilde{\alpha}\sqrt{R^4+
4\tilde{\alpha}R^{5-D}M}} - \frac{R}{2\tilde{\alpha}} <
0\label{eq:16a}
\end{eqnarray}
That is, the collapsing of the shell $r=const$ from the finite
initial data $R(0,r)=r$ is accelerated. Thus, $R$ will vanish
in finite proper time for the comoving observer, the shell focusing
singularity is inevitably formed in the three cases $W(r)=1$, $W(r)>1$
and $W(r)<1$. Furthermore, using Eq.~(\ref{eq:14a}) we obtain
\begin{eqnarray}
\frac{{d\dot{R}}}{{d\alpha }}
=\frac{\frac{2R^{5-D}M}{\sqrt{R^4+4\tilde{\alpha}R^{5-D}M}}
-\frac{\sqrt{R^4+4\tilde{\alpha}R^{5-D}M}-R^2}{\tilde{\alpha}}}{4\tilde{\alpha}\dot{R}}
= \frac{-P(Q-1)^2}{8{\tilde{\alpha}}^2 \dot{R}},\label{eq:17a}
\end{eqnarray}
where $P = ( R^4 + 4\tilde{\alpha}R^{5-D}M)^{\frac{1}{2}}$ and
$Q = \frac{R^2}{P}$. Apparently we have $\frac{d\dot{R}}{d\tilde{\alpha}}>0$,
and then the velocity of collapsing decreases with increasing $\tilde{\alpha} $.
This result indicates that positive $\tilde{\alpha}$ delays the
formation of the shell crossing singularity for $W(r)=1$,
$W(r)>1$ and $W(r)<1$ in higher dimensional $D\geq6$ space-times.

In the homogeneous collapse, the
metric Eq.~(\ref{eq:4a}) takes the form as the Robertson-Walker
metric \cite{Robertson:1935, Walker:1936}
\begin{eqnarray}
ds^2=- dt^2+a(t)^2(\frac{dr^2}{1+kr^2}+r^2d\Omega_{D-2}^2),\label{eq:18a}
\end{eqnarray}
where $k$ is a constant.
In order to satisfies Eq.~(\ref{eq:8a}), we can normalize the
radial coordinate $r$ so that $a( 0 ) = 1$. It is clear that the
time when the shell hits  shell focusing singularity is
completely determined by $a( t )=0$. Therefore it is
independent on $r$, as in the general relativity case. We can check such a consequence
by using Eq.~(\ref{eq:14a}). In the homogeneous case, $M( r )$ is
written in the form
\begin{eqnarray}
M( r ) = \frac{2}{(D-2)(D-1)}\varepsilon(0)r^{D-1}.\label{eq:19a}
\end{eqnarray}
Considering Eq.~(\ref{eq:18a}), Eq.~(\ref{eq:19a}) and
Eq.~(\ref{eq:14a}), we get
\begin{eqnarray}
\dot{a}(t)=-\sqrt{\frac{(a(t)^4+\frac{8\tilde{\alpha}\varepsilon(0)}{(D-1)(D-2)a(t)^{D-5}}
)^{\frac{1}{2}}-a(t)^2}{2\tilde{\alpha} }+k}.\label{eq:20a}
\end{eqnarray}
This equation does not contain the variable $r$, that is, $a( t )$
is a function of $t$. This implies that every shell
collapse into the shell focusing singularity at the same time.

Maeda proved that in the homogeneous collapse with the marginally bound case
$W(r)=1$, the shell crossing singularity is ingoing null for $D=5$
and is space-like for $D>5$ \cite{Maeda:2006pm}.
More generally, we
can extend such a consequence to cases $W(r)>1$ and $W(r)<1$.
Considering Eq.~(\ref{eq:20a}) we have
\begin{eqnarray}
\frac{2\varepsilon(0)}{(D-1)(D-2)}= \tilde{\alpha}(\dot{a}^2-
k)^2a^{D-5} + (\dot{a}^2- k)a^{D-3}.\label{eq:21a}
\end{eqnarray}
Since the energy density $\varepsilon ( 0 )$ have a finite
non-negative value, we find that the factor $a$
behaves as $c_1(t_{SF})(t-t_{SF})^{\frac{4}{D-1}}$ near the
singularity ${t = {t_{SF}}}$. We take the line element of the FRW
solutions with the factor $a =c_1(t_{SF})(t-t_{SF})^{\frac{4}{D-1}}$
to the conformally flat form as
\begin{eqnarray}
ds^2 = a^2(t(\tilde{t}, \tilde{r}))b^2(r(\tilde{r}))(- d\tilde{t}^2
+ d\tilde{r}^2+ \tilde{r}^2d\Omega_5^2),\label{eq:22a}
\end{eqnarray}
where
\begin{eqnarray}
d\tilde{t} &=& \frac{dt}{a(t)b(r)},\quad
\tilde{r} = \frac{r}{b(r)} \nonumber\\
b( r ) &=& \exp (\ln | r | + \ln | \frac{1}{r} + \sqrt
{\frac{1}{{{r^2}}} + k}  |).\nonumber
\end{eqnarray}
The range of $\tilde{t} $ for $t \in ( { - \infty ,{t_{SF}}} )$ is
$( { -\infty , + \infty } )$ and $(-\infty , \tilde{t}_0 )$,
for $D=5$ and $D>5$, respectively, where $\tilde{t}_0$ is a
constant. Thus, the shell focusing singularity is ingoing null for
$D=5$, while it is space-like for $D>5$, as shown in FIG.1.

Next, we consider whether such a shell focusing singularity is naked
or covered by an event horizon. An important constructions in
general relativity is that of the apparent horizon,
which is the outermost marginally trapped
surface for the outgoing null geodesics \cite{Hayward:1993wb}. The
apparent horizon is local in time and
can be located at a given space-like hypersurface, while the
event horizon coincides in case of static or stationary space-time
and is non-local. Moreover, the event horizon is a outer covering
surface of apparent horizons, that is, without the presence of
apparent horizons there is no event horizon. The condition for the existence of the apparent horizon with
outward normals null is
\begin{eqnarray}
{g^{\mu \nu }}{R_{,\mu }}{R_{,\nu }} =  - {\dot{R}^2} +
\frac{{{R'^2}}}{{{A^2}}} = 0.\label{eq:23a}
\end{eqnarray}
Using Eq.~(\ref{eq:6a}) and Eq.~(\ref{eq:13a}), the apparent horizon
condition becomes
\begin{eqnarray}
\frac{(R^4+4\tilde{\alpha}MR^{5-D})^{\frac{1}{2}}-
R^2}{2\tilde{\alpha}} + W^2-1 = W^2,\label{eq:24a}
\end{eqnarray}
then we obtain
\begin{eqnarray}
R_{AH}(t_{AH}(r),r)=\sqrt{\frac{M(r)}{R_{AH}^{D-5}} -
\tilde{\alpha}}.\label{eq:25a}
\end{eqnarray}
Clearly, the coupling constant $\tilde{\alpha}$ produces a change in
the location of these horizons.
Jhingan and Ghosh found that, in 5D case the positive $\tilde{\alpha}$ forbids
apparent horizon from reaching the coordinate center thereby making
the singularity massive and eternally visible, which is forbidden in
the corresponding general relativistic scenario
\cite{Christodoulou:1984mz, Cooperstock:1996yv}. We will show, in the $D>5$
space-time, the naked singularity would not be formed except at the
coordinate center. It is easy to find the condition that
Eq.~(\ref{eq:25a}) could not be satisfied before the formation of
the singularity is $M(r)<\tilde{\alpha}$,
in the 5D space-time, and the apparent horizon hits the singularity
when $M(r)=\tilde{\alpha}$.
In these cases, the shell focusing singularity can be visible from
infinity with certain initial data of energy density and leave open
even the weak form of CCH \cite{Christodoulou:1984mz}. Furthermore,
the shell focusing singularity is forever visible from future null
infinity after it has been formed if $M(r_b)<\tilde{\alpha}$
in the 5D case, where $r_b$ is the boundary of the collapsing dust
cloud. In other words, there is no black hole can be formed if the
mass function $M(r_b)$ takes sufficiently small value.

However, it can be seen obviously that in the $D\geq6$ space-time the apparent horizon lies in front of the
singularity unless $r=0$, thus the solution forbids the formation of
the naked singularity except the singular point at the center, as
same as in the general relativity \cite{Christodoulou:1984mz}.

For the homogeneous collapse, it is easy to find
\begin{eqnarray}
a(t_{AH}( r ))=\frac{R_{AH}}{r}=
\frac{1}{r}\sqrt{\frac{2}{(D-1)(D-2)}\varepsilon r^{4}a(t_{AH}( r ))^{5-D} -
\tilde{\alpha}}.\label{eq:26a}
\end{eqnarray}
Obviously, $a'( {{t_{AH}}( r )} ) > 0$. Since $\dot{a}( t )$ is negative,
the shell with greater coordinate value $r$ reaches ${R_{AH}}(
{{t_{AH}}( r ),r} )$ earlier, hence the singular shells which hit
the singularity at the same time could not be visible from infinity
with the presence of apparent horizons. As demonstrated above, our solution can be attached to the outside vacuum solution at the boundary of the dust cloud. Hence, using the conclusions of the global structure of space-time mentioned in \cite{Torii:2005xu}, we get that in the outside vacuum region, the singularity is outgoing time-like if
it is eternally visible, is outgoing space-like if the apparent
horizon lies backward it, and is outgoing null if the apparent
horizon hits it. We can give Penrose diagrams for the homogeneous
collapse as shown in FIG.1.

\begin{figure}
\centering \subfigure[$D\geq6$]{
\label{fig:subfig:A} 
\includegraphics{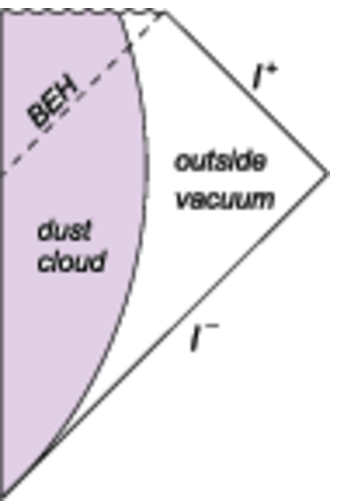}}%
\hfill%
\subfigure[$D=5, M(r)>\tilde{\alpha}$]{
\label{fig:subfig:B} 
\includegraphics{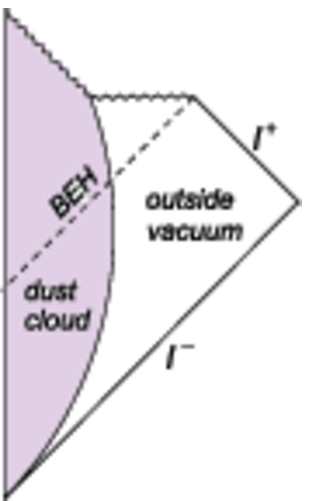}}
\hfill%
\subfigure[$D=5, M(r)=\tilde{\alpha}$]{
\label{fig:subfig:C} 
\includegraphics{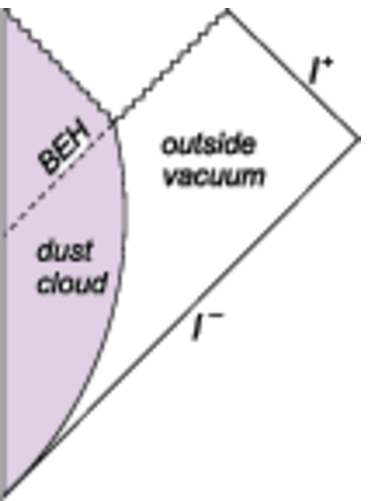}}
\hfill%
\subfigure[$D=5, M(r)<\tilde{\alpha}$]{
\label{fig:subfig:B} 
\includegraphics{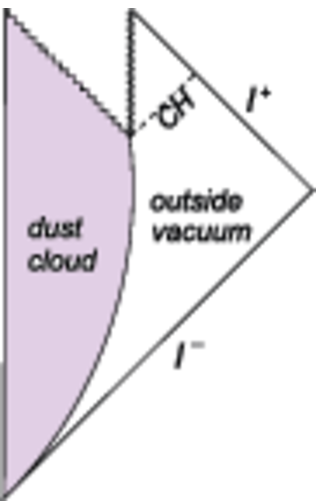}}
\caption{ Penrose diagram of the homogeneous collapse of a
spherically symmetric dust cloud in EGB. Zigzag lines represent the shell
focusing singularities, $I^{+(-)}$ corresponds to the future (past)
null infinity, BEH and CH stand for the black hole event horizon and
the Cauchy horizon, respectively. }
\label{fig:subfig} 
\end{figure}

In addition, it can be noted that Eq.~(\ref{eq:24a}) takes the same
form for cases $W(r)>1$, $W(r)<1$ and $W(r)=1$, therefore the discussion about
the apparent horizon in the case $W(r)=1$ which is given by Jhingan
and Ghosh can be applied to cases $W(r)>1$ and $W(r)<1$
\cite{Jhingan:2010zz}.

\section{Naked singularity and CCH\label{55s}}
\indent

In this section, we consider whether CCH is violated by naked singularities formed in the gravitational collapse in EGB. First we discuss the new CCH propose by Virbhadra, Ellis, Keeton, etc. This group classified naked singularities into 3 categories by their gravitational lensing effects: weakly naked, marginally strongly naked, and strongly naked  singularities. Naked singularities
contained within at least one photon sphere are termed
weakly naked singularities. Marginally strongly naked and strongly naked singularities are not covered within any photon spheres, and strongly naked singularities can produce images of gravitational lensing with negative time delays \cite{Ellis:2002mz, Virbhadra:2008mz}. The new CCH is that only weakly naked singularities can be formed in the realistic gravitational collapse. The photon sphere equation obtained by them is applied to a general static spherically
symmetric metric. Since the naked singularities formed in $D\geq6$ EGB is visible for only one moment and the metric is not static in this moment, so we only consider observational classification of naked singularities which formed in 5D EGB.

For a static spherically symmetric space-time
\begin{eqnarray}
ds^2=-b(r)dt^2+a(r)dr^2+d(r)r^2d\Omega_{D-2}^2,\nonumber
\end{eqnarray}
the photon sphere equation take the form of \cite{Ellis:2002mz, Virbhadra:2008mz}
\begin{eqnarray}
2d(r)b(r)+rd'(r)b(r)-r b'(r)d(r)=0.\label{eq:000a}
\end{eqnarray}
Since our solution can be attached to the outside vacuum solution, after all shells collapse into the singularity, the space-time can be described by Eq.~(\ref{eq:00a}), and the photon equation Eq.~(\ref{eq:000a}) becomes
\begin{eqnarray}
2F(r')-r'\frac{d F(r')}{dr'}=0.\label{eq:010a}
\end{eqnarray}
The positive real solution of  Eq.~(\ref{eq:010a}) is
\begin{eqnarray}
r'=[\frac{8m}{3\Omega_3}(\frac{2m}{3\Omega_3}-\tilde{\alpha})]^{\frac{1}{4}}, \quad \frac{2m}{3\Omega_3}>\tilde{\alpha}\nonumber
\end{eqnarray}
 which is equivalent to that $M>\tilde{\alpha}$. Under such condition, however,  the shell focusing
 singularity is not naked in 5D EGB as shown in the Penrose diagram. Hence, the photon equation have no real solution while the naked singularity has been formed in 5D EGB.
 Consequently, naked, massive and uncentral singularities formed in 5D EGB are marginally strongly or strongly
naked singularities, which is a counter example of the new CCH suggested by Virbhadra \cite{Virbhadra:2009mz}.
Physically, the gravitation collapse is a dynamical procedure. One should apply the  dynamical photon sphere
equation given in \cite{Claudel:2000yi} to analyze  the collapse.  The  equation takes the form \cite{Claudel:2000yi}
\begin{eqnarray}
\frac{d^2x^1}{(dx^0)^2}=-\frac{1}{2}g^{\theta\theta}n^a\partial_ag_{\theta\theta}
(n^1-\frac{dx^1}{dx^0}n^0)g_{bc}\frac{dx^b}{dx^0}\frac{dx^c}{dx^0}+(\frac{dx^1}{dx^0}
\Gamma^0_{ab}-\Gamma^1_{ab})\frac{dx^a}{dx^0}\frac{dx^b}{dx^0},\nonumber
\end{eqnarray}
with $
n^0=\psi g_{1a}\frac{dx^a}{dx^0}; n^1=\psi g_{0a}\frac{dx^a}{dx^0}$ and $
\psi=(-det g)^{-\frac{1}{2}}(-g_{ab}\frac{dx^a}{dx^0}\frac{dx^b}{dx^0})^{-\frac{1}{2}}$.
Since our concern is the final fate of the collapse rather than the dynamic process, we are regretful to abandon the discussion of this equation.

Next, we consider the strength
of the singularity \cite{Tipler:1977zza}. The importance
of the strength of the singularity lies in the fact that even if a
naked singularity occurs, if it is gravitationally weak in some
suitable sense, it may not have any physical implications and it may
perhaps be removable by extending the space-time
\cite{Clarke:1994cw}. A sufficient condition for a strong curvature singularity is
that in the limit of approach to the singularity, we must have along
at least one causal geodesic $\gamma ( s )$ \cite{Deshingkar:1998cb, Clarke:1987},
\begin{eqnarray}
\mathop {\lim }\limits_{s \to {s_0}} {( {s - {s_0}} )^2}\psi =
\mathop {\lim }\limits_{s \to {s_0}} {( {s - {s_0}}
)^2}{R_{ab}}{V^a}{V^b} > 0,\label{eq:27a}
\end{eqnarray}
where ${V^a}$ is the tangent vector to the geodesic. The idea
captured here is that in the limit of approach to such a
singularity, the physical objects get crushed to a zero size, and so
the idea of extension of space-time through it would not make sense,
characterizing this to be a genuine space-time singularity.

Jhingan and Ghosh pointed out that in the 5D space-time the
central singularity which can be naked may
be gravitational weak \cite{Jhingan:2010zz}. We will prove that
although the Gauss-Bonnet term weakens the strength of singularity,
the central singularity in the case $D>5$ which can also be visible
is gravitational strong. We consider radial time-like causal
geodesics ${U^\mu} = \frac{{d{x^\mu}}}{{d\tau }}$ with the
marginally bound case $W(r)=1$, here the affine parameter $\tau $ is
the proper time along particle trajectories. According to such
definition, ${U^\mu}$ satisfies
\begin{eqnarray}
{U^\mu}{U_\mu}= -1,\nonumber
\end{eqnarray}
that is,
\begin{eqnarray}
- {( {\frac{{d{x^t}}}{{d\tau }}} )^2} + {R'^2}{(
{\frac{{d{x^r}}}{{d\tau }}} )^2} =  - 1.\label{eq:28a}
\end{eqnarray}
Using the geodesic equation, we have
\begin{eqnarray}
\frac{d^2x^t}{d\tau^2}+
\frac{1}{2}(R'^2)_{,t}(\frac{dx^r}{d\tau})^2=0.\label{eq:29a}
\end{eqnarray}
Substituting Eq.~(\ref{eq:28a}) into Eq.~(\ref{eq:29a}), we obtain
that radial time-like geodesics must satisfy
\begin{eqnarray}
\frac{{d{U^t}}}{{d\tau }} + \frac{{\dot{R}'}}{{R'}}[ {{{( {{U^t}}
)}^2} - 1} ] = 0.\label{eq:30a}
\end{eqnarray}
The simplest solution is the worldline of a freely falling particle,
which is ${U^\mu} = \frac{{d{x^\mu}}}{{d\tau }} = \delta _t^a$. In
terms of proper time we can describe it as
\begin{eqnarray}
{t_{SF}}( r ) - t = {\tau _0} - \tau .\label{eq:31a}
\end{eqnarray}
We consider the expansion of $\varepsilon(r)$ near $r=0$
\begin{eqnarray}
\varepsilon(r) = \sum\limits_{n = 0}^{ + \infty } {{\varepsilon_n}{r^n}}
\simeq {\varepsilon_0},\label{eq:32a}
\end{eqnarray}
it specifies that $R(t,r)$ behaves as ${R_0}( t )r$ near the coordinate center,
where ${R_0}( t )$ is a function of $t$ and vanishes at $t={t_{SF}}$.
Thus, we get
\begin{eqnarray}
\psi =R_{ab}U^aU^b= -\frac{(D-1)\ddot{R}_0}{R_0}.\label{eq:33a}
\end{eqnarray}
With the help of Eq.~(\ref{eq:16a}), we obtain that
\begin{eqnarray}
\mathop {\lim }\limits_{\tau \to \tau _0} (\tau -\tau
_0)^2\psi=\mathop {\lim }\limits_{\tau  \to \tau _0}(\tau-\tau _0
)^2[-(D-1)(\frac{\frac{1+(5-D)\tilde{\alpha}M_0R_0^{1-
D}}{\sqrt{1+4\tilde{\alpha}M_0R_0^{1-D}}} - 1}{2\tilde{\alpha}})],\label{eq:34a}
\end{eqnarray}
where $M_0$ is defined as $M_0= \mathop {\lim }\limits_{r \to 0}
\frac{{M( r )}}{{{r^{D-1}}}}$ therefore has finite value. In the
limit $t-t_{SF}\to 0$ we have
\begin{eqnarray}
\mathop {\lim }\limits_{\tau  \to {\tau _0}} \frac{{R_0(t)}}{{\tau -
{\tau _0}}}=\mathop {\lim }\limits_{t  \to {t _{SF}}}
\frac{{R_0(t)-R_0(t_{SF})}}{{t - {t _{SF}}}}=\mathop {\lim
}\limits_{t \to {t_{SF}}} {\dot{R}_0(t)},\nonumber
\end{eqnarray}
thus,
\begin{eqnarray}
\mathop {\lim }\limits_{\tau  \to {\tau _0}} {\left( {\tau  - {\tau
_0}} \right)^2} = \mathop {\lim }\limits_{\tau  \to {\tau _0}}
\frac{{2\tilde{\alpha} }}{{\sqrt {1 + 4\tilde{\alpha} M_0R_0^{1 -
D}} - 1}}.\label{eq:35a}
\end{eqnarray}
Consequently,
\begin{eqnarray}
\mathop {\lim }\limits_{\tau  \to {\tau _0}} ( {\tau  - {\tau _0}}
)^2\psi  = \frac{{(D-1)(D-5)}}{4},\label{eq:36a}
\end{eqnarray}
the strong curvature condition is satisfied in the case $D>5$ and is
not satisfied in the case $D=5$.

Hence, the central shell focusing singularity is gravitational strong in the $D>5$ space-time while it may be gravitational weak
in the 5D space-time. The
difference between two cases is that the singularity can be
eternally visible in the 5D case while it is disallowed in the $D>5$
case. Therefore, the singularity is gravitational strong when the
solution represents the black hole formation for arbitrary initial
data. If a naked singularity is gravitational weak, it may not have
any significant physical consequences and so may not be a serious threat
to the weak CCH of Penrose and the new CCH of Ellis's group. Thus, the naked singularity in the $D\geq6$ case
violates the CCH, as in the general relativity
\cite{Christodoulou:1984mz, Deshingkar:1998cb}, while the naked
singularity may not be regarded as an essential
counter example to the CCH seriously in the 5D case.

Furthermore, such a central singularity in the 5D case is gravitational strong in the
general relativistic limit. In the limit $\alpha \to 0$,
in $D\geq5$ space-time, Eq.~(\ref{eq:34a}) takes the form
\begin{eqnarray}
\mathop {\lim }\limits_{\tau  \to {\tau _0}} {( {\tau  - {\tau _0}}
)^2}\psi  = \mathop {\lim }\limits_{\tau  \to {\tau _0}} {( {\tau -
{\tau _0}} )^2}[ {\frac{{(D-1)(D-3){M_0}}}{{2R_0^{D-1}}}} ].\nonumber
\end{eqnarray}
and Eq.~(\ref{eq:35a}) becomes
\begin{eqnarray}
\mathop {\lim }\limits_{\tau  \to {\tau _0}} {\left( {\tau  - {\tau
_0}} \right)^2} = \mathop {\lim }\limits_{\tau  \to {\tau _0}}
\frac{{R_0^{D-1}}}{{{M_0}}}.\nonumber
\end{eqnarray}
Thus,
\begin{eqnarray}
\mathop {\lim }\limits_{\tau  \to {\tau _0}} ( {\tau  - {\tau _0}}
)^2\psi  =  \frac{(D-1)(D-3)}{2}  > 0,\label{eq:37a}
\end{eqnarray}
the strong curvature condition is satisfied in the 5D space-time.
Comparing Eq.~(\ref{eq:37a}) to Eq.~(\ref{eq:36a}),
we find that high order Lovelock terms weaken the strength of the singularity.

\section{conclusions and discussions \label{66s}}
\indent

In this paper, we have investigated the gravitational collapse of a
spherically symmetric dust cloud in the EGB
without a cosmological constant, and extended works of
Maeda, Jhingan and Ghosh to a more general case.
As we demonstrated, the final fate of the
collapsing dust cloud is a shell focusing singularity which can be
either hidden or naked. We analyzed the global structure of the
space-time and gave the Penrose diagram for the homogeneous case.
It turns out that, the contribution of Gauss-Bonnet curvature
corrections alters the course of
collapse and the time of formation of singularities,
modifies apparent horizon formation and the location of apparent
horizons, and changes the strength of singularities.
Unlike the 5D case, there is a
serious threat to the CCH caused by the naked, gravitational strong
singularity in the $D\geq6$ case.

In addition, it is interesting to
study the spherically symmetric gravitational collapse of a dust cloud
in higher order Lovelock gravity. We will discuss them elsewhere.

{\bf Acknowledgment }
This work has been supported by the Natural Science Foundation of China
under grant No. 10875060, 10975180 and 11047025.

\end{document}